\documentclass[]{article}
\usepackage{lmodern}
\usepackage{amssymb,amsmath}
\usepackage{ifxetex,ifluatex}
\usepackage{fixltx2e} 
\ifnum 0\ifxetex 1\fi\ifluatex 1\fi=0 
  \usepackage[T1]{fontenc}
  \usepackage[utf8]{inputenc}
\else 
  \ifxetex
    \usepackage{mathspec}
  \else
    \usepackage{fontspec}
  \fi
  \defaultfontfeatures{Ligatures=TeX,Scale=MatchLowercase}
\fi
\IfFileExists{upquote.sty}{\usepackage{upquote}}{}
\IfFileExists{microtype.sty}{%
\usepackage{microtype}
\UseMicrotypeSet[protrusion]{basicmath} 
}{}
\usepackage{hyperref}
\hypersetup{unicode=true,
            pdftitle={In oder Aus},
            pdfauthor={Ethan Madison; Zachary Zipper},
            pdfkeywords={bloom filter; set membership; genome assembly; cuckoo filter; quotient filter; fuzzy-folded; data structure},
            pdfborder={0 0 0},
            breaklinks=true}
\usepackage{longtable,booktabs}
\usepackage{graphicx,grffile}
\makeatletter
\def\maxwidth{\ifdim\Gin@nat@width>\linewidth\linewidth\else\Gin@nat@width\fi}
\def\maxheight{\ifdim\Gin@nat@height>\textheight\textheight\else\Gin@nat@height\fi}
\makeatother
\setkeys{Gin}{width=\maxwidth,height=\maxheight,keepaspectratio}
\ifx\paragraph\undefined\else
\let\oldparagraph\paragraph
\renewcommand{\paragraph}[1]{\oldparagraph{#1}\mbox{}}
\fi
\ifx\subparagraph\undefined\else
\let\oldsubparagraph\subparagraph
\renewcommand{\subparagraph}[1]{\oldsubparagraph{#1}\mbox{}}
\fi
\usepackage[binary-units]{siunitx}
\newcommand{\tta}{\texttt{A}}
\newcommand{\ttc}{\texttt{C}}
\newcommand{\ttg}{\texttt{G}}
\newcommand{\ttt}{\texttt{T}}
\providecommand{\subtitle}[1]{%
  \usepackage{titling}
  \posttitle{%
    \par\large#1\end{center}}
}

\title{In oder Aus}
\providecommand{\subtitle}[1]{}
\subtitle{Modern Improvements Upon and Applications of the Bloom Filter}
\author{Ethan Madison \and Zachary Zipper}
\date{\today}

\begin{document}
\maketitle
\begin{abstract}
Bloom filters are data structures used to determine set membership of
elements, with applications from string matching to networking and
security problems. These structures are favored because of their reduced
memory consumption and fast wallclock and asymptotic time bounds.

Generally, Bloom filters maintain constant membership query time, making
them very fast in their niche. However, they are limited in their lack
of a removal operation, as well as by their probabilistic nature. In
this paper, we discuss various iterations of and alternatives to the
generic Bloom filter that have been researched and implemented to
overcome their inherent limitations.

Bloom filters, especially when used in conjunction with other data
structures, are still powerful and efficient data structures; we further
discuss their use in industy and research to optimize resource
utilization.
\end{abstract}

\section{Introduction to Bloom
filters}\label{introduction-to-bloom-filters}

A Bloom filter is a probabilistic data structure used to test set
membership queries in constant time. Queries may return false positives,
but never a false negative (thus classifying them as a false-biased
Monte Carlo algorithm). Standard Bloom filters include insert and
set-membership query operations, and lack element removal, iteration,
and other features common in binary search trees, hash tables, or more
common data structures that can be used similarly.

\subsection{Construction of a standard Bloom
filter}\label{construction-of-a-standard-bloom-filter}

\begin{figure}[htbp]
\centering
\includegraphics{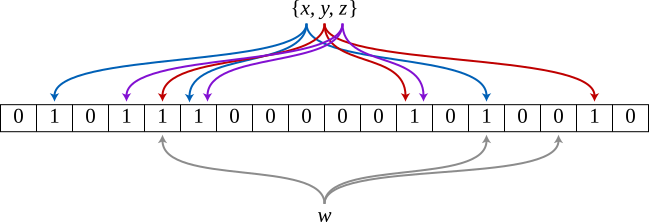}
\caption{An example Bloom filter storing the set \(\{x,y,z\}\) with
\(h=3\) hash functions. \(x\), \(y\), and \(z\) are each mapped by three
unique hash functions to three bits. \(w\) is queried by checking the
three bits that it hashes to. Since one of \(w\)'s corresponding bits is
unset, the query will return ``false.'' \label{fig:wxyz}}
\end{figure}

A Bloom filter represents a set of \(n\) items, and consists of \(h\)
unique hash functions and an array of \(m\) bits. To add an element to
the Bloom filter, compute its hash with each of the \(h\) functions and
set the bit at each index. (If the bit has been set previously, keep it
set.) To query an element, compute its hashes and return ``true'' if all
\(h\) bits are set; return false otherwise. Figure \ref{fig:wxyz}
demonstrates a query operation.

An element \(f\) not in the set could exist such that all \(h\) of its
corresponding bits have been set by other elements. Querying \(f\) would
return ``true'': a \emph{false positive}. A query will never mistakenly
report that an element is not a member of the set when it actually
is---in other words, there are no \emph{false negatives}.

\subsection{False positives}\label{false-positives}

\begin{figure}[htbp]
\centering
\includegraphics{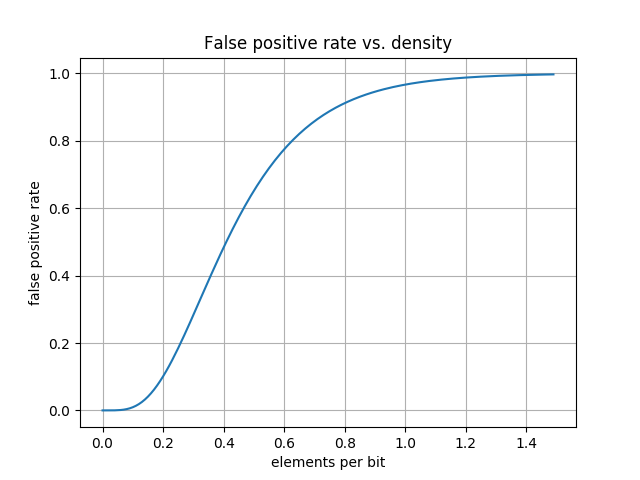}
\caption{A plot of Equation \ref{eqn:fpr}: false positive rate vs.
\(\frac{n}{m}\), or the number of elements per bit of the Bloom filter,
when \(h=5\). Note that Bloom filter false positive rates are ideally at
most .02, where the rate of change is approximately 0---i.e., the first
``few'' (relative to \(m\)) elements inserted negligibly affect the
false positive rate. \label{fig:fpr-plot}}
\end{figure}

The false positive rate of Bloom filters can be estimated {[}1{]} using

\begin{equation}\text{FPR} = \left(1 - \left(1-\frac{1}{m}\right)^{hn}\right)^h \approx \left(1 - e^{-\frac{hn}{m}}\right)^h,\label{eqn:fpr}\end{equation}

where \(m\) is the size of the Bloom filter, \(h\) is the number of hash
functions, and \(n\) is the number of elements inserted. The approximate
form is graphed in Figure \ref{fig:fpr-plot}.

To fix the false positive rate at \(\text{FPR}\), the optimal number of
hash functions to use {[}2{]} is

\begin{equation}h = \log_2 \left(\frac{1}{\text{FPR}}\right).\end{equation}

\subsection{Example use case}\label{example-use-case}

Bloom filters are often used to reduce memory or storage requirements of
set membership data structures. An example problem in which Bloom
filters could be used is browser vendors protecting users from malicious
websites: while browser vendors could ship a list of known malicious
URLs, shipping the full list would require excessively large network
transaction and storage space. Instead, the browser vendor could include
in the browser package a Bloom filter containing that list, and browsers
could query the local Bloom filter quickly before loading any webpage.
If the query returns ``true'', then the webpage can be checked via the
Internet against the exact list.\footnote{In fact, this is how Google
  Chrome protects users from malicious sites. As of 2010, the full list
  contained \textasciitilde{}1 million websites, stored in an only 18 Mb
  Bloom filter {[}3{]} {[}4{]}.} If the query returns ``false'', then
the webpage is not on the list---a query for an element actually on the
list will never return ``false''.

\subsection{Theoretic advantages}\label{theoretic-advantages}

The primary advantage of Bloom filters is how they markedly beat other
data structures used for set membership queries in terms of memory
overhead. Naive and deterministic implementations of set membership data
structures (including binary search trees and hash tables) generally
have to store the entirety of the elements that they represent. A
significant improvement, Bloom filters represent an element with one bit
per \(h\) hash functions. (Typically, fewer than 10 bits are required to
represent an element in a Bloom filter {[}5{]}.) Furthermore, the number
of elements in the filter does not need to correlate with the number of
bits (or the size of the array) used to store that element.

There are some extensions to the common Bloom filter that make this
advantage even more extreme (notably the Fuzzy-Folded Bloom filter,
which can support \textasciitilde{}1.9 times the elements of a standard
Bloom filter while maintaining the same false positive rate and a
constant time complexity in all operations).

A secondary advantage of Bloom filters is their ability to insert and
query elements in constant time. Binary search trees can do neither in
constant time. Hash tables can often only query elements in amortized
constant time, and generally have a linear worst case for insertion. The
latencies of Bloom filter operations are solely dependent on the number
and complexities of the hash functions it employs.

Because of Bloom filters' ability to quickly process elements with low
space cost, they are particularly useful in problems that involve data
streams {[}6{]} (or massive sets of elements with no defined upper
bounds on size). For example, a massive hash table that is used to test
set membership will eventually fill up and break when a boundless number
of elements are inserted. While the load factor of a Bloom filter in the
same scenario may become dangerously high, the core functionality of the
data structure will always remain intact, even when inserting an
extremely large number of elements.

\subsection{Limitations}\label{limitations}

The probabilistic nature of Bloom filters may make them unsuitable for
certain tasks---though that can be mitigated by choosing an appropriate
size of the bit array and number of hash functions to achieve
sufficiently low false positive rates. By Equation \ref{eqn:fpr}, one
can increase the size \(m\) and manipulate (depending on
\(\frac{n}{m}\)) the number of hash functions \(h\) to decrease the
false positive rate to an acceptable percentage. To eliminate false
positives, the Bloom filter can be used as a preliminary check to
eliminate negatives before checking against the exact list. (This is
faster than checking everything against the exact list, since Bloom
filter queries are in constant time and Bloom filters are small enough
to fit in faster, limited-quantity memory (e.g., cache).)

The inability to remove elements from a Bloom filter makes the structure
unfit for highly dynamic and volatile sets, where the membership of
elements changes rapidly. (Even re-inserting elements into a new Bloom
filter doesn't work, since a Bloom filter cannot efficiently and
precisely report the entire set; the hashing of elements and subsequent
setting of bits is an irreversible operation.) However, there are
several extensions on the common Bloom filter which add removal
support---these will be discussed in the next section.

\section{Developments}\label{developments}

In this section, we will discuss a few iterative improvements on the
standard Bloom filter, as well as the dynamic Fuzzy-folded Bloom filter
and the ``practically better'' Cuckoo filter. These improvements include
addition of a removal operation, superior space complexity, and better
hardware interaction in the interest of practicality. While all of these
solve the same general sorts of problems, one may be better than another
for specific use cases. Table \ref{table:filters} compares several
properties of the discussed filters.

\subsection{Counting Bloom filter}\label{counting-bloom-filter}

The most primitive Bloom filter iteration supporting element removal is
the Counting Bloom filter. In this variation, each bucket or bit is
replaced by a number of bits (usually four), which serve as a counter.
If an inserted element hashes to an index, that index is incremented. If
an element is removed, then all the counters it hashes to are
decremented. When querying an element, the query returns ``true'' if all
the counters it hashes to are greater than 0; the query returns false
otherwise. Disadvantages of this variation relative to the standard
Bloom filter include the possibility of bit overflow (in the counters)
and higher memory overhead (as it requires storing several bits for each
array index instead of only one).

\subsection{Blocked Bloom filter}\label{blocked-bloom-filter}

Blocked Bloom filters (like standard Bloom filters) do not support
removal. However, they are constructed to exhibit high spatial locality
by fitting several smaller Bloom filters individually into cache lines.
They guarantee less than two cache misses on negative queries {[}7{]},
where standard Bloom filters can have up to \(h\) misses on negative
queries. Thus, a Blocked Bloom filter can be a much more practical data
structure than a general Bloom filter. However, for these improvements,
implementing a Blocked Bloom filter requires some knowledge of the
hardware.

\subsection{\texorpdfstring{\(d\)-left Counting Bloom
filter}{d-left Counting Bloom filter}}\label{d-left-counting-bloom-filter}

\(d\)-left Counting Bloom filters use fingerprinting to support removal.
They insert elements by computing their \(d\)-left hash to store as a
fingerprint---removal is accomplished simply by deleting the computed
fingerprint. In this way, they are almost more similar to a standard
hash table than a bloom filter. This structure's space cost is up to
twice as expensive as that of a standard Bloom filter, but it requires
half the space of a Counting Bloom filter {[}5{]}. The idea of
fingerprinting incorporated in this structure is also a concept
fundamental to Quotient and \protect\hyperlink{cuckoo}{Cuckoo filters}.

\subsection{Quotient filter}\label{quotient-filter}

Quotient filters are another extension of the standard Bloom filter
which also use string fingerprinting to support removal. The main
difference between Quotient filters and \(d\)-left Counting Bloom
filters is how Quotient filters use an extension on standard hashing
called ``quotienting'' to efficiently hash and reconstruct the
fingerprints of elements. In quotienting, the high and low bits of a
computed fingerprint are partitioned; the low bits are then stored in a
bucket indexed by the high bits. Due to this scheme, collisions
generally display high spatial locality and therefore are
hardware-friendly {[}8{]}. However, this structure still relies on
linear probing in order to resolve collisions. Thus, Quotient filters
cannot give the same guarantees on asymptotic time complexity as most
other Bloom filter implementations, and its performance suffers at
\(\geq 75\%\) load {[}2{]}.

\begin{longtable}[c]{@{}llll@{}}
\caption{Characteristics of Bloom filters, extensions, and variations.
The cache misses column shows the worst-case. In the General Bloom and
the Blocked Bloom designs, \(h\) indicates the number of hash functions
used. In the \(d\)-Left Counting Bloom construction, \(d\) is the number
of partitions in its hash table {[}2{]}. As explained in the following
section, \(f\) is the number of completed
fuzzy-folds.\label{table:filters}}\tabularnewline
\toprule
Filter type & Space cost & Cache misses per lookup & Deletion
support\tabularnewline
\midrule
\endfirsthead
\toprule
Filter type & Space cost & Cache misses per lookup & Deletion
support\tabularnewline
\midrule
\endhead
General Bloom & 1\(\times\) & \(h\) & No\tabularnewline
Counting Bloom & \(3\times \sim 4\times\) & \(h\) & Yes\tabularnewline
Blocked Bloom & \(1\times\) & 1 & No\tabularnewline
\(d\)-Left Counting Bloom & \(1.5\times \sim 2\times\) & \(d\) &
Yes\tabularnewline
Quotient & \(1\times \sim 1.2\times\) & \(\geq 1\) & Yes\tabularnewline
Fuzzy-Folded Bloom & \(\sim 0.5\times\) & \(h(f+2)\) & No\tabularnewline
Cuckoo & \(\leq 1\times\) & 2 & Yes\tabularnewline
\bottomrule
\end{longtable}

\subsection{Fuzzy-folded Bloom filter}\label{fuzzy-folded-bloom-filter}

A Fuzzy-folded Bloom filter describes the continuous compression
(``folding'') of two standard Bloom filters (created from a bipartition
of the original array) each of size \(\frac{m}{2}\) bits into a single,
compressed filter with \(\frac{m}{2}\) buckets. Exactly half of the
space of the original array is allocated to this compressed Bloom
filter, and the other half is used to support two new Bloom filters each
of size \(\frac{m}{4}\) bits {[}6{]}.

In this, the Fuzzy-folded Bloom filter is a ``dynamic'' Bloom filter,
growing to maintain a low false positive rate while accommodating
further insertions. The fuzzy-folding operation does not break the
invariants of the original filter (in that it will never introduce the
possibility of false negatives), nor does it increase the false positive
rate of the filter {[}6{]}.

The fuzzy-folding operation overlays bits at the same position in both
Bloom filters and uses fuzzy logic in each bucket of the product array
to represent the compression of these filters. This process (and
therefore compressed bit representation) is non-commutative; it is
necessary to logically reconstruct the ordering to effectively query the
filter {[}6{]}.

Insertion is similar to standard Bloom filter insertion. There are
always two non-compressed Bloom filters in the array. There is also a
designated load threshold, applicable to both non-compressed filters. If
the first array has yet to reach this threshold, \(h\) bits are set in
the first array (whose indices are determined by the element's \(h\)
hashes). If the first array has reached this threshold, \(h\) bits are
set in the second array. If both arrays have reached this threshold,
then they are fuzzy-folded, and two new Bloom filters are allocated with
\(m/\left(2^{2+f}\right)\) bits each (where \(f\) denotes the number of
fuzzy-fold operations conducted before this operation, starting from 0).

Queries operate by first checking the second non-compressed filter (and
returning ``true'' when a standard Bloom filter would). If there is no
match, the first non-compressed filter is then checked. If there is
again no match, then the filters are sequentially queried from most
recently to least recently compressed. Therefore, in the worst case, the
time complexity of queries is linear in terms of the number of
fuzzy-folds performed. Also, in all cases, the speed of queries suffers
drastically compared to that of a standard Bloom filter when the size of
the compressed arrays becomes sufficiently small {[}6{]}.

The most notable advantage of Fuzzy-folded Bloom filters is how they can
accommodate roughly 1.9 times the elements of a general, space-optimized
Bloom filter while maintaining the same false positive rate {[}6{]}. In
practice, this is a desirable trade-off, considering linear query
complexity in terms of number of fuzzy-folds is not a significant
limitation, or even comparable to linear in terms of inserted elements.

\subsection{Iterative Patterns}\label{iterative-patterns}

Bloom filter evolutions are built with practicality in mind: due to the
real-world performance boost from cache optimization, many of these
filters are constructed for high spatial locality. Likewise, many
datasets are volatile, and so fingerprinting is commonly used to let a
Bloom filter variant support removal without introducing unacceptable
space overhead.

\hypertarget{cuckoo}{\subsection{Cuckoo Filter}\label{cuckoo}}

Cuckoo filters use many of the same paradigms and ideas of Bloom
filters, fill the same niche (of performing highly space and time
efficient set membership queries with no false negatives), yet approach
a few key concepts in ways that differentiate them from Bloom filters.

\begin{figure}[htbp]
\centering
\includegraphics{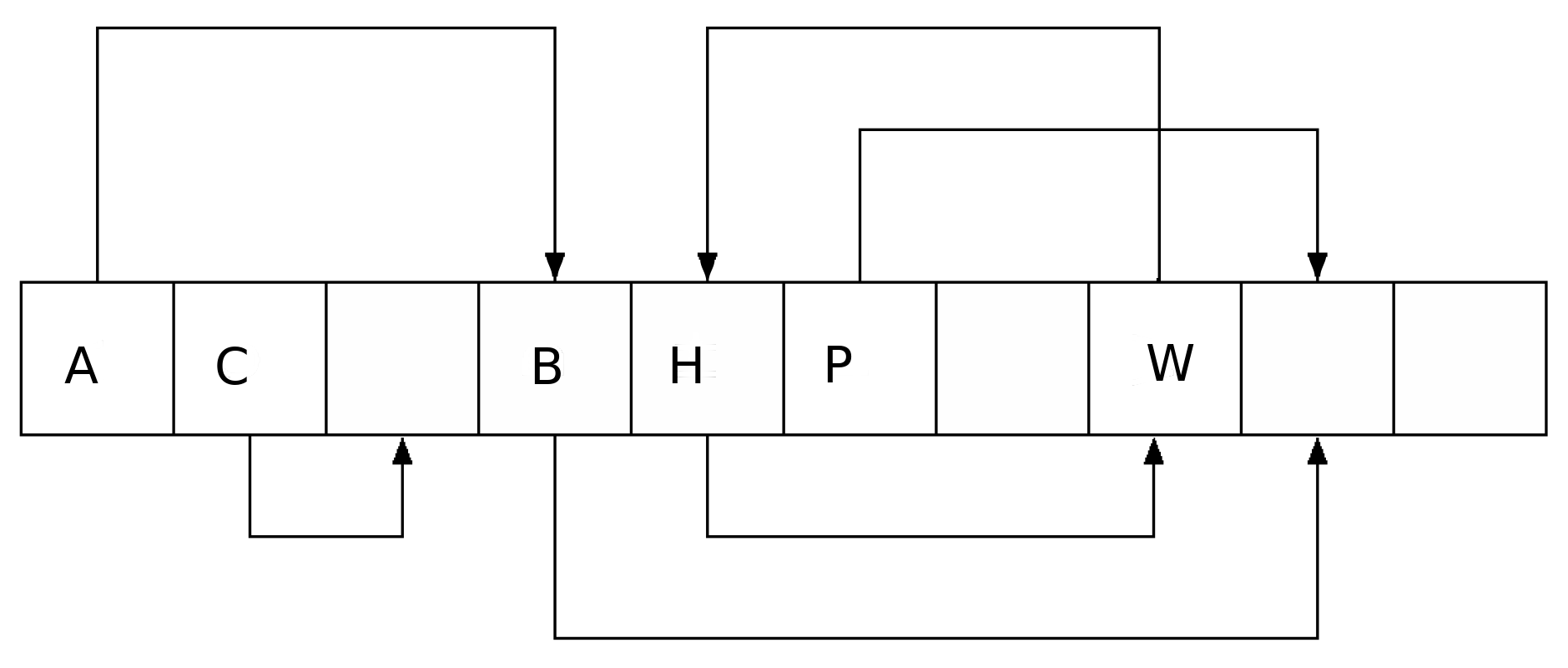}
\caption[\label{fig:cuckoo}An illustration of the Cuckoo hashing
technique used by Cuckoo filters. Each element has two possible buckets.
Inserting into an occupied bucket displaces the occupying element into
its alternate bucket. E.g., inserting an element that first hashes to
where \(C\) is would cause the new element to displace \(C\) to \(C\)'s
alternate bucket. Inserting an element that first hashes to where \(A\)
is would displace \(A\) into the bucket \(B\) currently occupies,
displacing \(B\) into its empty alternate bucket. Inserting an element
that first hashes to the bucket \(W\) would cause an infinite loop since
\(W\) and \(H\) hash to the same two buckets.]{\label{fig:cuckoo}An
illustration of the Cuckoo hashing technique used by Cuckoo filters.
Each element has two possible buckets. Inserting into an occupied bucket
displaces the occupying element into its alternate bucket. E.g.,
inserting an element that first hashes to where \(C\) is would cause the
new element to displace \(C\) to \(C\)'s alternate bucket. Inserting an
element that first hashes to where \(A\) is would displace \(A\) into
the bucket \(B\) currently occupies, displacing \(B\) into its empty
alternate bucket. Inserting an element that first hashes to the bucket
\(W\) would cause an infinite loop since \(W\) and \(H\) hash to the
same two buckets.\footnotemark{}}
\end{figure}
\footnotetext{Original figure by Ramus Pagh, via
  \href{https://en.wikipedia.org/wiki/Cuckoo_hashing\#/media/File:Cuckoo.svg}{Wikipedia},
  under a
  \href{https://creativecommons.org/licenses/by-sa/3.0/deed.en}{Creative
  Commons License (CC BY-SA 3.0)}. Figure has been rotated.}

Notably, Cuckoo filters support element removal. They accomplish this by
using a fingerprinting scheme similar to that seen in \(d\)-left
Counting Bloom filters and Quotient filters, where deleting an element
amounts to deleting its fingerprint. Specifically, removing an element
from a Cuckoo filter is accomplished by checking both of the element's
possible buckets and deleting the fingerprint if it exists in either
{[}2{]}.

A Cuckoo filter is also more capable of taking advantage of its
allocated space. Because of Cuckoo filters' insertion scheme
(illustrated in Figure \ref{fig:cuckoo}), which sets a fingerprint in
only one (compared to \(h\)) bucket per insertion and (in some
implementations) allows buckets to contain multiple fingerprints, a load
factor of \(95\%\) without a marked increase in false positives is very
achievable {[}2{]}.

The final primary advantage of Cuckoo filters is in their simple design
and construction. Compared to more complicated variations on Bloom
filters that reduce time and space complexity at comparable margins,
such as Blocked Bloom filters and Golomb-Compressed Sequences {[}2{]},
the concepts and implementation details behind Cuckoo filters are
relatively simple.

On the other hand, Cuckoo filter insertion is arguably worse than that
of a standard Bloom filter. The process of insertion has the same worst
case as general Cuckoo hashing, where all buckets for the hashed
fingerprint are occupied, leading to a chain reaction of displacements
throughout the entire table. Nevertheless, Cuckoo filters still maintain
amortized \(\mathcal{O}(1)\) insertion {[}2{]}.

Cuckoo filters also have an upper bound on the number of times one
fingerprint can be inserted: if the Cuckoo filter has buckets of size
\(b\), then elements with the same fingerprint can be inserted at most
\(2b\) times {[}2{]}. Disabling the removal operation can overcome this
limitation, but removal is a sought-after feature in this niche.

To maintain an acceptable false-positive rate, fingerprint size must
scale with the size of the filter (or its number of buckets).
Nevertheless, this is generally acceptable, as Cuckoo filters are more
space efficient than the standard Bloom filter at low false positive
rates (\textasciitilde{}\(3\%\)) {[}2{]}.

\section{Bloom Filters in Practice}\label{bloom-filters-in-practice}

We give an in-depth discussion of a new \emph{de novo} genome assembler,
ABySS 2.0, which utilizes Bloom filters to trivialize memory
requirements without sacrificing speed or accuracy, and increases DNA
sequencing throughput. This and further advancements could revolutionize
preventative healthcare, and affect the lives of many. We continue to
list many of the ways Bloom filters are used in everyday life to make
efficient otherwise difficult procedures.

\subsection{\texorpdfstring{\emph{De novo} genome
assembly}{De novo genome assembly}}\label{de-novo-genome-assembly}

Genomics research---the field of characterizing genomes to better
understand similarities and differences among species, or even
individuals---has seen much development in the past decade in part
thanks to growing and planned personalized medicine
initiatives.\footnote{``Personalized medicine'' is a medical model which
  tailors medical decisions, practices, interventions and/or product
  usage to the individual patient's predicted risks. Because of this
  focus on the individual, and given the large part genetics play in
  individual health, the field depends on sequencing genomes \emph{en
  masse}. The spread of personalized medicine may be a major boon to
  preventative healthcare, especially as sequencing becomes faster and
  more affordable.} During this time, the DNA sequence throughput of the
industry's best instruments has constantly increased {[}1{]}. In
particular, sequence assembly, which refers to aligning and merging
fragments read or copied from a longer DNA sequence in order to
reconstruct the original sequence,\footnote{DNA sequencing technology,
  biological or synthetic, cannot read whole genomes (on the order of 7
  billion basepairs split among 23 pairs of chromosomes in humans) in
  one pass. Instead, genome sequencing works by copying or listing the
  bases in a short (20--30000 bases) ``reads'' and then combining them.
  \emph{De novo} sequence assembly, which constructs genomes without a
  backbone or template, combines these reads using overlaps as
  indicators of originally adjacent sequences. Figure
  \ref{fig:Seqassemble} illustrates the \emph{de novo} assembly process.}
has seen drastic reductions in both time and spatial requirements with
the incorporation of Bloom filters and related algorithms in assemblers.

\begin{figure}[htbp]
\centering
\includegraphics{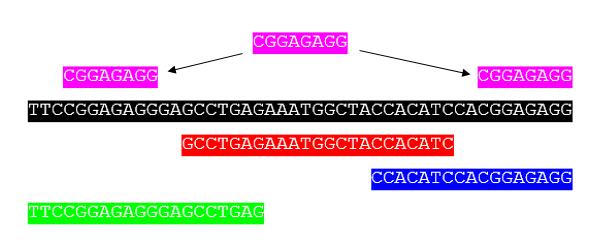}
\caption[Sample sequence showing how a sequence assembler would take
short fragments (pink, red, blue, and green) and match by overlaps to
recreate the black sequence. Notice that the pink fragment could be in
one of two positions in the sequence. \label{fig:Seqassemble}]{Sample
sequence showing how a sequence assembler would take short fragments
(pink, red, blue, and green) and match by overlaps to recreate the black
sequence. Notice that the pink fragment could be in one of two positions
in the sequence.\footnotemark{} \label{fig:Seqassemble}}
\end{figure}
\footnotetext{Figure (presented without modification) by ``Luongdl'',
  via
  \href{https://en.wikipedia.org/wiki/Sequence_assembly\#/media/File:Seqassemble.png}{Wikipedia},
  under a
  \href{https://creativecommons.org/licenses/by-sa/3.0/}{Creative
  Commons License (CC BY-SA 3.0)}.}

The genome assembler ABySS is example of the efficiency of Bloom filters
in \emph{de novo} genome assembly: whereas ABySS 1.9 (which didn't use a
Bloom filter) could assemble the human genome in \SI{14}{\hour} using a
whopping \SI{418}{\giga\byte} of memory (across many machines), ABySS
2.0 can, using the same parameters, assemble the human genome in
\SI{20}{\hour} with a mere \SI{34}{\giga\byte}\footnote{The absolute
  time for sequencing doesn't matter: this is about throughput. This is
  an order of magnitude increase in throughput given the same resources,
  which matters for real-world applications. Roughly 10 times the number
  of patients can have their DNA sequenced using this sort of algorithm.}
{[}1{]}. ABySS 2.0 achieves this performance increase by consolidating
its usage to a single machine (eliminating the need for messaging) and
instead represents a De Bruijn graph using a Bloom filter. In this
context, a De Bruijn graph stores all \(4^k\) possible length-\(k\)
sequences (called ``\(k\)-mers'', which are length \(k\) portions of the
longer read) made up of symbols (bases) from
\(\{\tta, \ttc, \ttg, \ttt\}\) in the vertices
\[V = \left\{ (\tta, \dots, \tta, \tta), (\tta, \dots, \tta, \ttc), \dots, (\tta, \dots, \tta, \ttt), (\tta, \dots, \ttc, \tta), \dots, (\ttt, \dots, \ttt, \ttt) \right\}\]
and all four possible ``successor'' sequences (where the first base is
removed, the rest shifted left, and another base is appended) are
represented in the edges
\[E = \left\{((v_1, v_2, \dots, v_n), (v_2, \dots, v_n, s_i) : i=1, \dots, m)\right\}.\]

\begin{figure}[htbp]
\centering
\includegraphics{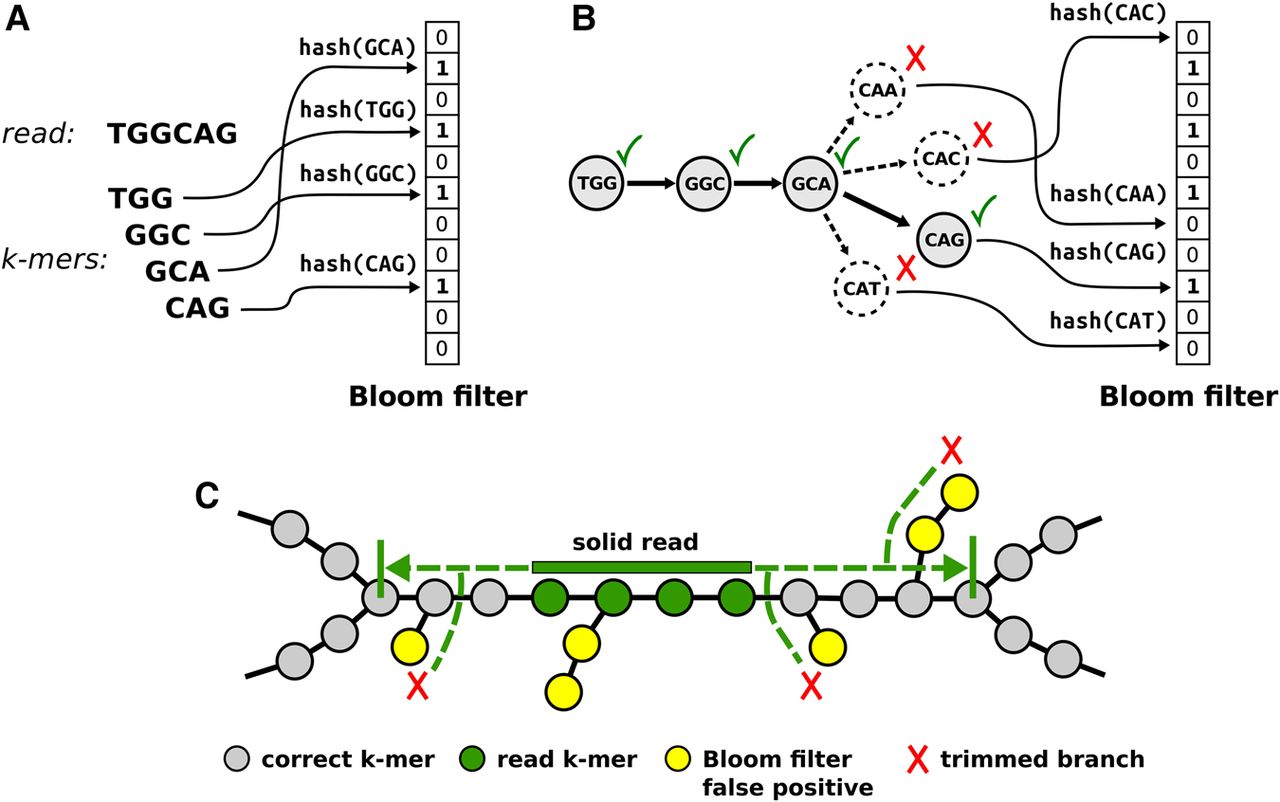}
\caption[\label{fig:abyss2-algo}An overview of the ABySS 2.0 assembly
algorithm. \textbf{A}: A read is split into \(k\)-mers and loaded into
the Bloom filter by computing the hash of each \(k\)-mer sequence and
setting the corresponding bit in the Bloom filter. \textbf{B}: A path
through the De Bruijn graph is traversed by querying all possible
successor \(k\)-mers and advancing to those found. \textbf{C}: ABySS 2.0
trims dead-end branches and continues only along those at least a fixed
length. The term ``solid read'' denotes a confirmed sequence. Edges are
discovered in both directions: both predecssors and succesors are
searched for {[}1{]}.]{\label{fig:abyss2-algo}An overview of the ABySS
2.0 assembly algorithm. \textbf{A}: A read is split into \(k\)-mers and
loaded into the Bloom filter by computing the hash of each \(k\)-mer
sequence and setting the corresponding bit in the Bloom filter.
\textbf{B}: A path through the De Bruijn graph is traversed by querying
all possible successor \(k\)-mers and advancing to those found.
\textbf{C}: ABySS 2.0 trims dead-end branches and continues only along
those at least a fixed length. The term ``solid read'' denotes a
confirmed sequence. Edges are discovered in both directions: both
predecssors and succesors are searched for {[}1{]}.\footnotemark{}}
\end{figure}
\footnotetext{Figure (presented without modification) from {[}1{]},
  under a \href{https://creativecommons.org/licenses/by/4.0/}{Creative
  Commons License (Attribution 4.0 International license)}}

As described in Figure \ref{fig:abyss2-algo}, ABySS 2.0 represents the
vertices of the De Bruijn graph with a Bloom filter by setting bits
corresponding to the vertices (\(k\)-mers). These \(k\)-mers are the
length-\(k\) sequences of a short read, so in this way the Bloom filter
contains the entire read. With all reads stored in the Bloom filter,
ABySS 2.0 combines them to reconstruct the original sequence. The
algorithm repeatedly queries the Bloom filter to discover De Bruijn
graph edges (which lead to predecessor or successor \(k\)-mers). Since
\(k\)-mers are relatively short and are only part of the read, this
process may lead to representation of false edges in the graph;
fortunately, using look-ahead mechanisms, these branches are trimmed if
they do not continue for more than \(k\) nodes. (This look-ahead
mechanism increases graph traversal cost, but eliminates the need for
additional data structures.) As reads extend and more likely represent a
correct path in the De Bruijn graph, they are branded ``solid reads''
and considered to be part of the original sequence. Through this
process, the entire genome may be reconstructed from short reads with
high accuracy.\footnote{The accuracy is ``on-par with other assemblers''
  {[}1{]}, and need not be perfect; even the biological process of DNA
  transcription introduces error. Additionally, DNA has some protections
  against error in that many codons encode the same meaning, as well as
  the fact that much of the genome has little-to-no known effect.}

Table \ref{table:assembler-comp} shows how ABySS 2.0 compares to other
genome assembly algorithms in resource consumption. Notably, the tools
which efficiently represent the De Bruijn graph (ABySS 2.0, MEGAHIT,
Minia, and SGA) require much less memory than those that do not.
BCALM\footnote{BCALM uses a novel method of partitioning the De Bruijn
  graph to be resource efficient, and does not use Bloom filters.},
while extremely resource light, sacrifices sequence contiguity compared
to the rest of of the listed assemblers. ABySS 2.0, using Bloom filters,
achieves a marked improvement over ABySS 1.9 and the rest of the pack in
efficiency, while achieving contiguity results on par with
DISCOVARdenovo and ABySS 1.9. Note that Minia also uses Bloom filters,
and ABySS 2.0 is largely based on Minia, with three novel features: (i)
the use of solid reads, (ii) a look-ahead mechanism to eliminate false
positives (as opposed to a seperate data structure), and (iii) a new
hashing algorithm designed for DNA/RNA sequences. The authors of ABySS
believe that there is still great opportunity for improving throughput
without sacrificing contiguity---i.e., the algorithmic ideas of BCALM
could be adapted to produce a more contiguous result {[}1{]}. (And
perhaps Bloom filters are not the absolute most efficient means of
representing a De Bruijn graph.)

\begin{longtable}[c]{@{}lll@{}}
\caption{Peak memory usage and wallclock runtime with 64 threads of
assemblies of GIAB HG004.\label{table:assembler-comp} Data from
{[}1{]}.}\tabularnewline
\toprule
Assembler & Memory (GB) & Time (h)\tabularnewline
\midrule
\endfirsthead
\toprule
Assembler & Memory (GB) & Time (h)\tabularnewline
\midrule
\endhead
ABySS 1.9 & 418 & 14\tabularnewline
ABySS 2.0 & 34 & 20\tabularnewline
DISCOVARdenovo & 618 & 26\tabularnewline
BCALM & 5 & 9\tabularnewline
MEGAHIT & 197 & 26\tabularnewline
Minia & 137 & 19\tabularnewline
SGA & 82 & 65\tabularnewline
SOAPdenovo & 659 & 35\tabularnewline
\bottomrule
\end{longtable}

\subsection{Networking}\label{networking}

Bloom filters are used in device discovery: if two previously paired
devices meet again under different circumstances, they can skip pairing
again (which would be unnecessary, since they have previously done so).
Devices build a Bloom filter of devices they have paired with, and when
attempt to connect with another device, send the list to the second
device. If the second device recognizes one of its identifiers in the
Bloom filter, it responds to the first device that they have paired
previously, which initiates the mutual connection. This process,
described and patented in {[}9{]}, is used in Qualcomm devices (i.e.,
many or most cell phones and other portable devices) for ad-hoc network
discovery (e.g., Bluetooth, WiFi direct, 802.xx wireless LAN).

\subsection{Making the world go 'round}\label{making-the-world-go-round}

Facebook {[}10{]} uses Bloom filters to represent the social graph for
typeahead search in order to display friends and friends-of-friends of
the user's query. The Bloom filter uses 16 bits per friend connection
(or graph edge).

Yahoo! Mail {[}11{]} uses Bloom filters to represent email contact list
since the Bloom filter can fit in browser cache. This obviates the need
for round-trip connections to Yahoo for verifying delivered emails are
from contacts.

Tinder {[}12{]} {[}11{]} uses Bloom filters to record ``right swipes''
(accepting a user as a possible match) in order to remove users from the
incoming list. When eventually the list is refreshed for new users to
swipe on, those previously right swiped on will be filtered out. Some
unseen users are filtered out by this process, but, as the saying goes,
there are plenty of fish in the sea.

URL shorteners {[}11{]} employ Bloom filters to generate unique URLs: if
a shortened URL has been previously used, it exists in the Bloom filter.
Thus, by querying the Bloom filter with different shortened URLs until
receiving a ``false'' response, the service can ensure unique URLs.

YouTube uses Bloom filters {[}11{]} to ensure recommended videos are not
in the user's watch history, in addition to the algorithms that optimize
for relevance metrics, to feed users new and interesting content.

\section{Conclusion}\label{conclusion}

The general Bloom filter is a powerful data structure thanks to its
simplicity and its time and memory efficiency. Guaranteed constant time
insert and query operations are rare among data structures, and often
worth the trade-off of false positives in set membership problems. A
more serious limitation for particular problems is the lack of a removal
operation, which is overcome by various evolutions of the Bloom filter,
and the Cuckoo filter. For many problems, one of these evolutions or
alternatives may be better suited depending on the circumstantial
resources and constraints.

Bloom filters, though simple and often hidden behind-the-scenes, have a
profound and increasing effect on the modern world. This data structure
makes efficient dating, entertainment, and networking---both social and
digital---possible, as well as leading healthcare to be able to
personalize treatment to the individual.

\newpage

\setlength\parindent{0pt}

\section*{References}\label{references}
\addcontentsline{toc}{section}{References}

\hypertarget{refs}{}
\hypertarget{ref-abyss2}{}
{[}1{]} S. D. Jackman, ``ABySS 2.0: Resource-efficient assembly of large
genomes using a bloom filter,'' \emph{Genome Research}, vol. 27, no. 5,
pp. 768--777, 2017.

\hypertarget{ref-cuckoo}{}
{[}2{]} B. Fan, D. G. Andersen, M. Kaminsky, and M. D. Mitzenmacher,
``Cuckoo filter: Practically better than bloom,'' \emph{Proceedings of
the 10th ACM International on Conference on emerging Networking
Experiments and Technologies - CoNEXT 14}, pp. 75--88, Dec 2014.

\hypertarget{ref-yakuninux5f2010}{}
{[}3{]} A. Yakunin, ``Nice bloom filter application,'' \emph{Alex
Yakunin's blog}. Mar-2010.

\hypertarget{ref-hessux5f2012}{}
{[}4{]} S. Hess, ``Issue 10896048: Transition safe browsing from bloom
filter to prefix set. - code review,'' \emph{Appspot.com}. Jan-2012.

\hypertarget{ref-bonomiux5f2006}{}
{[}5{]} F. Bonomi, M. Mitzenmacher, R. Panigrahy, S. Singh, and G.
Varghese, ``An improved construction for counting bloom filters,''
\emph{Lecture Notes in Computer Science Algorithms -- ESA 2006}, pp.
684--695, 2006.

\hypertarget{ref-fuzzy}{}
{[}6{]} A. Singh, S. Garg, K. Kaur, S. Batra, N. Kumar, and K.-K. R.
Choo, ``Fuzzy-folded bloom filter-as-a-service for big data storage on
cloud,'' \emph{IEEE Transactions on Industrial Informatics}, pp. 1--1,
Jun 2018.

\hypertarget{ref-putzeux5f2009}{}
{[}7{]} F. Putze, P. Sanders, and J. Singler, ``Cache-, hash-, and
space-efficient bloom filters,'' \emph{Journal of Experimental
Algorithmics}, vol. 14, p. 4.4, Dec 2009.

\hypertarget{ref-benderux5f2012}{}
{[}8{]} M. A. Bender, ``Don't thrash,'' \emph{Proceedings of the VLDB
Endowment}, vol. 5, no. 11, pp. 1627--1637, Jul 2012.

\hypertarget{ref-qualcommux5f2015}{}
{[}9{]} W. Haddad, M. Vandereen, G. Tsirtsis, and V. D. Park, ``Bloom
filter based device discovery.'' Nov-2015.

\hypertarget{ref-adamsux5f2010}{}
{[}10{]} K. Adams, ``Typehead search tech talk,'' \emph{Facebook.com}.
Facebook Engineering, Jan-2010.

\hypertarget{ref-quora}{}
{[}11{]} ``What are the best applications of bloom filters? -
quora\_2014,'' \emph{Quora.com}. Jan-2014.

\hypertarget{ref-senux5f2018}{}
{[}12{]} G. Sen, ``Designing tinder: System design interview question,''
\emph{YouTube}. Jul-2018.

\end{document}